\documentclass[letterpaper,12pt]{article}
\usepackage{osajnl2}
\usepackage{graphicx}
\begin{document}

\title
{ Wavelengths of the $\rm \bf 3d^6(^5D)4s\,a^6D - 3d^6\,(^5D)4p\,y^6P $
Multiplet of Fe II (UV 8)}

\author{Gillian Nave$^*$, Craig J. Sansonetti}
\address{ National Institute of Standards and Technology, Gaithersburg, 
MD~20899, U.S.A. }
\address{$^*$Corresponding author: gnave@nist.gov}

\begin{abstract}

We investigate the wavenumber scale of Fe~I and Fe~II lines using new
spectra recorded with Fourier transform spectroscopy and using a
re-analysis of archival spectra. We find that standards in Ar~II,
Mg~I, Mg~II and Ge~I give a consistent wavenumber calibration. We use
the recalibrated spectra to derive accurate wavelengths for the
a$^6$D-y$^6$P multiplet of Fe~II (UV 8) using both directly measured
lines and Ritz wavelengths. Lines from this multiplet are important
for astronomical tests of the invariance of the fine structure
constant on a cosmological time scale. We recommend a wavelength of
1608.45081~\AA\ with a one standard deviation uncertainty of
0.00007~\AA\ for the $\rm a^6D_{9/2}-y^6P_{7/2}$ transition.

\end{abstract}
\maketitle

\section{Introduction}\label{intro}

The universality and constancy of the laws of nature rely on the
invariance of the fundamental constants. However,
some recent measurements of quasar (Quasi-stellar objects - QSO)
absorption line spectra suggest that the fine-structure constant,
$\alpha$, \cite{fund_const} may have had a different value during
the early universe \cite{murphy_03}. Other measurements
(e.g. \cite{chand_06}) do not show any change. The attempt to resolve these
discrepancies can probe deviations from the standard model of particle
physics and thus provide tests of modern theories of fundamental
interactions that are hard to attain in other ways.

QSO absorption lines are used in these investigations 
to measure the wavelength separations of atomic lines in spectra of different
elements - the many-multiplet method \cite{Dzuba_99} - and compare
their values at large redshifts with their values today. Any
difference in the separations would suggest a change in $\alpha$. Since
this method uses many different species in the analysis that have differing
sensitivities to changes in $\alpha$, it can be much more sensitive 
than previous methods, such as the alkali-doublet
method \cite{Bahcall_67}, that use just one species.
However, it requires very accurate laboratory wavelengths to be used
successfully, since the observed changes in $\alpha$ are only a few
parts in 10$^5$, requiring laboratory wavelengths to 1:10$^7$ or
better. This has led to several recent measurements of
ultraviolet wavelengths using both Fourier transform (FT) spectroscopy
\cite{Pickering_98,Aldenius_06,Aldenius_09} and frequency comb metrology
\cite{Salumbides_06,Hannemann_06,Batteiger_09}.

One spectral line of particular interest is the $\rm
3d^6(^5D)4s\,a^6D_{9/2} - 3d^5\,(^6S)4s4p(^3P)\,y^6P_{7/2} $ line of Fe~II at
1608.45~\AA. This line is prominent in many QSO spectra and its
variation with $\alpha$ has the opposite sign from that of other
nearby lines \cite{Murphy_03}. However, measurement of its wavelength
using frequency comb metrology, which is at present the most accurate
method, is extremely difficult due to its short wavelength. Although
this line is strong in many of the FT spectra of iron-neon hollow
cathode lamps recorded at the National Institute of Standards and
Technology (NIST) and Imperial College, London, UK (IC), these
spectra display inconsistencies in the wavelength of the 1608~\AA\
line of around 1.5 parts in 10$^7$ - too great for use
in the many-multiplet method to detect changes in $\alpha$. We
discussed some of these discrepancies in our previous paper
\cite{Nave_04} presenting reference wavelengths in the spectra of
iron, germanium and platinum around 1935~\AA . 

Here we present a re-analysis of spectra taken at NIST and IC in order
to resolve these discrepancies and provide a better value for the
wavelength of the 1608.45~\AA\ line of Fe~II. The papers involved in
this re-analysis are listed in table \ref{corrections}, together with 
the proposed corrections to the wavenumber scale. The proposed corrections
are up to three times the previous total uncertainty, depending on the
wavenumber. In section
\ref{previous} we discuss previous measurements of the  a$^6$D -
y$^6$P multiplet. Section \ref{expt} describes the archival data we
use to obtain improved wavelengths for this multiplet Additional
spectra taken at NIST in order to re-evaluate the calibration of these
archival data are described in section \ref{cal}. The accuracy of this
calibration in the visible and ultraviolet wavelength regions is also
discussed in section \ref{cal}. Section \ref{a6D-y6P} describes three
different methods for obtaining the wavelengths of the a$^6$D - y$^6$P
multiplet. The first method uses intermediate levels determined using
strong Fe~II lines in the visible and ultraviolet in order to obtain
the values of the y$^6$P levels and Ritz wavenumbers for the a$^6$D -
y$^6$P multiplet. The second method uses energy levels optimized by
using a large number of spectral lines to derive Ritz wavenumbers for
this multiplet.  Although better accuracy is achieved using this method than the first
method because of the increased redundancy, the way in which the
y$^6$P levels are determined is less transparent. The third method
uses experimental wavelengths determined in spectra that are
recalibrated from spectra in which we have re-evaluated the wavenumber
calibration. In section \ref{ge_comp} we re-examine the Fe~II
wavenumbers in our previous paper \cite{Nave_04}. All uncertainties in
this paper are reported at the one standard deviation level.

\section{Previous measurements of the a$^6$D - y$^6$P multiplet}\label{previous}

\begin{figure}
\includegraphics[width=130mm]{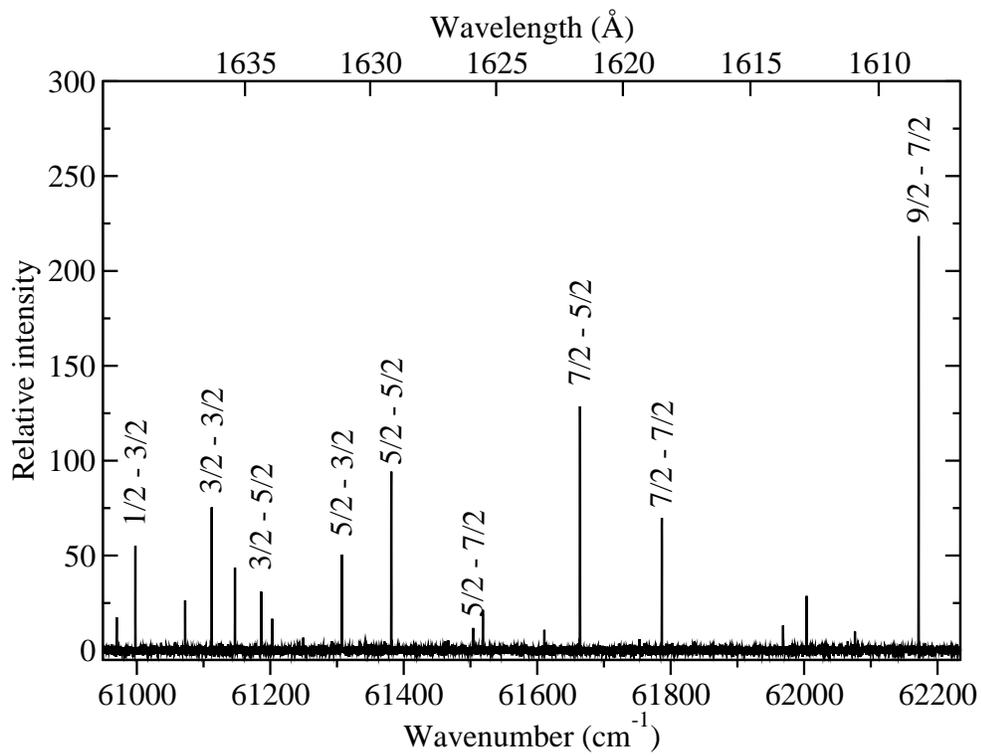}
\caption{The region of the Fe~II $\rm a^6D - y^6P$ transitions. The labeled
lines show the J-values of the lower and upper energy levels
respectively.
\label{1608_plot}}
\end{figure}

The region of the a$^6$D - y$^6$P multiplet is shown in figure
\ref{1608_plot} as observed in a FT spectrum taken at IC.
Nave, Johansson \& Thorne \cite{Nave_97} report Ritz and
experimental wavelengths for six of the nine
lines of the a$^6$D - y$^6$P multiplet. The Ritz wavelengths are based
on energy levels optimized to 
spectral lines covering wavelengths from 1500~\AA\ to 5.5~$\rm \mu$m
measured with FT spectroscopy. The estimated uncertainties are about 2x10$^{-4}$~\AA\
or about 1.2 parts in 10$^7$. The published lines do not include the
$\rm a^6D_{9/2} - y^6P_{7/2} $ line. Johansson
\cite{Johansson_78} contains Ritz wavelengths for all nine lines,
based on unpublished interferometric measurements of
Norl\'{e}n, with a value of 1608.451~\AA\ for this line. The estimated uncertainty
of the $\rm y^6P_{7/2}$ level with respect to the ground state  $\rm
a^6D_{9/2}$  of Fe~II  is 0.02~cm$^{-1}$. The uncertainty of
the 1608~\AA\ line can be derived directly from the uncertainty of the
$\rm y^6P_{7/2}$ level, and corresponds to a
wavelength uncertainty at 1608~\AA\ of 0.0005~\AA . Wavelengths for
all nine lines measured using FT spectroscopy are also given in
Pickering et al. \cite{Pickering_02} in a paper devoted to oscillator
strength measurements. No details of the calibration of these lines or
their uncertainties are given. The wavelength value
recommended by Murphy et al. \cite{Murphy_03} is
1608.45080$\pm$0.00008~\AA , with a reference to Pickering et
al. However, this is not the value given in Pickering et al. and 
the small uncertainty is improbable without additional confirmation. The
source of this wavelength is unclear.

In addition to these published values, lines from this multiplet are
present in some unpublished archival spectra from IC and NIST. The
most important spectra for the current work are summarized in Table
\ref{spectra_summary}. The spectra on which Nave, Johansson \& Thorne
\cite{Nave_97} is based are part of a much larger set of Fe~II spectra
covering all wavelengths from 900~\AA\ to 5.5~$\rm \mu$m. Two of these
spectra cover the region around 1600~\AA\ and contain all nine lines
of the multiplet. The wavelength standards for these spectra are
traceable to a set of Ar~II lines between 3729~\AA\ and 5146~\AA\ (see
section \ref{cal} for details).
The weighted average wavelength for the $\rm a^6D_{9/2} - y^6P_{7/2} $
line in these unpublished archival spectra is 1608.45075$\pm$0.00018~\AA .

The spectra in Nave \& Sansonetti \cite{Nave_04} were calibrated with
respect to Ge standards of Kaufman \& Andrew \cite{Kaufman_62}. In
addition to the spectra used in that paper, we recorded a spectrum
using FT spectroscopy 
with a pure iron cathode that covers the wavelength region of the $\rm
a^6D - y^6P$ multiplet (fe1115 in Table
\ref{spectra_summary}). It was calibrated with iron lines measured in
one of the spectra used for ref. \cite{Nave_04} (lp0301 in Table
\ref{spectra_summary}). The resulting value for the wavelength of
the $\rm a^6D_{9/2} - y^6P_{7/2} $ line was
1608.45050$\pm$0.00004~\AA, 1.5x10$^{-7}$ times smaller than the 
wavelength obtained from the archival spectra and outside their joint
uncertainty. This inconsistency is also larger than the uncertainty
required for measurements of possible changes in $\alpha$.

\section{Summary of current experimental data}\label{expt}

The spectra we re-analyzed are the same as those used in previous
studies of Fe~I and Fe~II
\cite{Learner_88,Nave_91,Nave_97, Nave_04}. Three
different spectrometers were used: the f/60 IR-visible-UV FT
spectrometer at the National Solar Observatory, Kitt Peak, AZ (NSO),
the f/25 vacuum ultraviolet (VUV) FT spectrometer at IC
\cite{Thorne_87}, and the f/25 VUV spectrometer at NIST
\cite{Griesmann_99}. The light sources for all of the spectra were
high-current hollow cathode lamps containing a cathode of pure iron run
in either neon or argon. Gas pressures of 100~Pa to 500~Pa (0.8~Torr
to 4~Torr) were used with currents from 0.32~A to 1~A. The total
number of FT spectra was 31, covering wavelengths from about 1500~\AA\
to 5~$\rm \mu$m (2000~cm$^{-1}$ to 66000~cm$^{-1}$). The wavenumber,
intensity and width for all the lines were obtained with Brault's {\sc
decomp} program \cite{Brault_89} or its modification {\sc xgremlin}
\cite{Nave_97b}. Further details of the experiments can be found in
\cite{Learner_88,Nave_91,Nave_97,Nave_04}.  Additional spectra were
taken using the NIST 2-m FT spectrometer and are described in section
\ref{vis_NIST}. 

\section{Calibration of FT spectra}\label{cal}

All of the spectra were calibrated assuming a linear FT wavenumber scale,
so that in principle only one reference line is required to
put the measurements on an absolute scale. In practice, many lines are
used. To obtain the absolute wavenumbers, a multiplicative correction 
factor, k$_{\mbox{eff}}$, is derived from the reference lines and applied
to each observed wavenumber $\sigma_{\mbox{obs}}$ so that
\begin{equation}
\sigma_{\mbox{corr}} = (1 + k_{\mbox{eff}})\sigma_{\mbox{obs}}
\end{equation}
where $\sigma_{\mbox{corr}}$ is the corrected wavenumber.

All the spectra in Learner \& Thorne (3830~\AA\ to 5760~\AA )
\cite{Learner_88} and Nave et al. (1830~\AA\ to 3850~\AA )
\cite{Nave_91} trace their calibration to 28 Ar~II lines in the
visible. The original calibration in Refs. \cite{Learner_88} and
\cite{Nave_91} used the wavenumbers of Norl\'{e}n \cite{Norlen_73} for these
lines. Norl\'{e}n calibrated these Ar~II lines with respect to
$^{86}$Kr~I lines emitted from an electrodeless microwave discharge
lamp that had in turn been calibrated with respect to an Engelhard
lamp, which was the prescribed source for the primary wavelength
standard at the time of his measurements. The estimated standard
uncertainty of Norl\'{e}n's Ar~II wavenumbers varies from
0.0007~cm$^{-1}$ at 19429~cm$^{-1}$ to 0.001~cm$^{-1}$ at
22992~cm$^{-1}$. The Ar~II lines were used to calibrate a `master
spectrum' (spectrum k19 in Table \ref{spectra_summary}). Additional
spectra of both Fe-Ne and Fe-Ar hollow cathode lamps covering
wavelengths from 2778~\AA\ to 7387~\AA\ were calibrated from this
master spectrum.   

The ultraviolet spectra reported in \cite{Nave_91} were calibrated
with respect to the results of Learner \& Thorne \cite{Learner_88} by
using a bridging spectrum. This bridging spectrum used two different
detectors, one on each output of the FT spectrometer. The first overlapped
with the visible wavenumbers in Ref. \cite{Learner_88} in order to
obtain a wavenumber calibration and the second covered the UV
wavenumbers being measured.  Since the two outputs of the FT
spectrometer are not exactly in antiphase, the resulting phase
correction has a discontinuity in the region around 35~000~cm$^{-1}$
where the two detectors overlap, as shown in Fig. 1 of
\cite{Nave_91}. The full procedure is described in detail in
\cite{Nave_91}.

The 28 Ar~II lines used as wavenumber standards in
Refs. \cite{Learner_88,Nave_91} were subsequently re-measured by
Whaling et al. \cite{Whaling_95} using FT spectroscopy with molecular
CO lines as standards. The uncertainty of these measurements is
0.0002~cm$^{-1}$. The molecular CO standards used in
Ref. \cite{Whaling_95} were measured using heterodyne frequency
spectroscopy with an uncertainty of around 1:10$^9$ and are ultimately
traceable to the cesium primary standard \cite{Maki_92}. The
wavenumbers of Whaling et al. \cite{Whaling_95} are systematically
higher than those of Norl\'{e}n \cite{Norlen_73} by 6.7$\pm$0.8 parts in
10$^8$, corresponding to a wavenumber discrepancy  of about
0.0014~cm$^{-1}$ at 21000~cm$^{-1}$. Since the results of Whaling et
al. \cite{Whaling_95} are more accurate and precise than those of
Norl\'{e}n \cite{Norlen_73}, all the wavenumbers in \cite{Learner_88},
\cite{Nave_91}, and Table 3 of \cite{Nave_97} have been increased by
6.7 parts in 10$^8$ wherever they are used in the current work. 

The spectra in Nave \& Sansonetti \cite{Nave_04} were calibrated with
respect to 29 Ge~I Ritz wavenumbers derived from the energy levels of
Kaufman \& Andrew \cite{Kaufman_62}. However, the Fe~II wavenumbers
derived using this calibration were found to be greater than those in
Nave et al. \cite{Nave_91} by about 7 parts in 10$^8$, even after the
wavenumbers in the latter were adjusted to the wavenumber scale of
Whaling et al. \cite{Whaling_95}. 

In order to present accurate wavenumbers for Fe~II lines around
1600~\AA, it is necessary first to confirm the accuracy of the iron
lines in the visible that were calibrated with respect to selected
lines of Ar~II lines \cite{Learner_88}, to investigate the accuracy with which this
calibration is transferred to the VUV, and to resolve the discrepancy
between iron and germanium standard wavelengths identified in Ref. \cite{Nave_04}.

\subsection{Calibration of the visible-region spectra}\label{vis_NIST}

In order to confirm the calibration of the master spectrum, k19, used
in \cite{Learner_88} and \cite{Nave_91}, we took additional spectra
using the NIST 2~m FT spectrometer \cite{Nave_97b}. The source was a
water-cooled high-current hollow cathode lamp with a current of 1.5~A
and argon at pressures of 130~Pa to 330~Pa (1 Torr to 2.5 Torr). The
spectra covered the region 8500~cm$^{-1}$ to 37~000~cm$^{-1}$ with
resolutions of either 0.02~cm$^{-1}$ or 0.03~cm$^{-1}$.  A 1~mm
aperture was used in order to minimize possible illumination
effects. The detector was a silicon photodiode detector with a
2~mm~$\times~$2~mm active area.

The spectrometer was aligned optimally using a diffused, expanded beam
from a helium neon laser, ensuring that the modulation of the laser
fringes was maximized throughout the 2~m scan. Before recording some
of the spectra, the spectrometer was deliberately misaligned and
re-aligned in order to test whether small misalignments that could not
be detected using our alignment procedure affected the wavenumber
scale.

The spectra were calibrated using the values of Whaling et al.
\cite{Whaling_95} for Ar~II lines recommended in Ref. \cite{Learner_88} that had good
signal-to-noise ratio. Wavenumbers of strong iron lines were then
measured and compared with iron lines taken from Ref.
\cite{Learner_88} and \cite{Nave_91}.

\begin{figure}
\includegraphics[width=130mm]{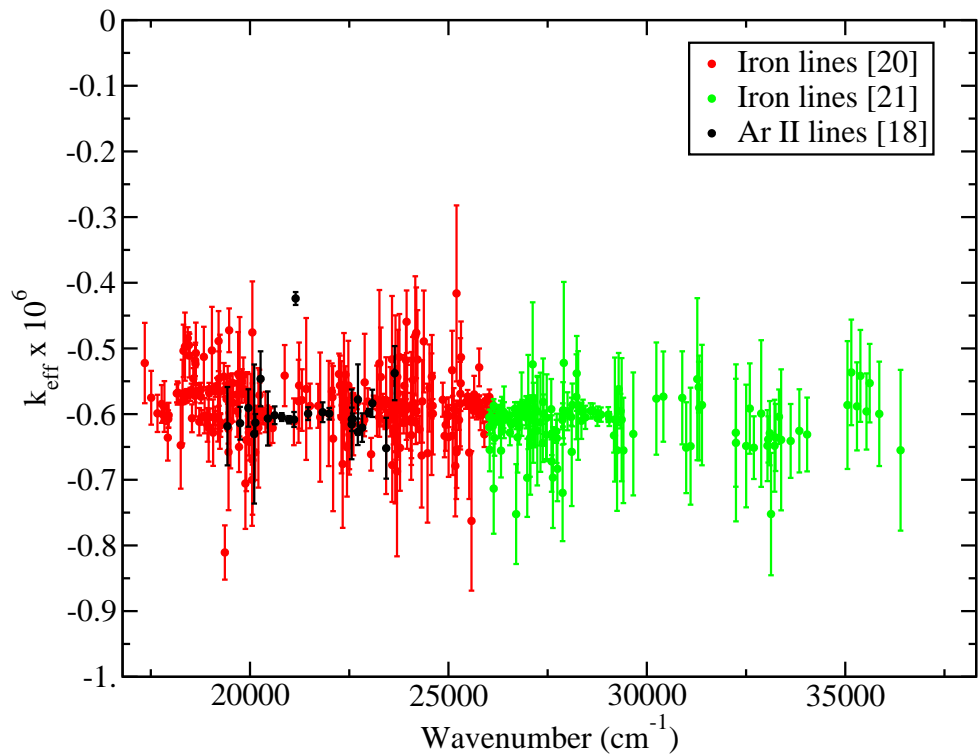}
\caption{ (Color online) Calibration of wavenumbers in spectrum 
fe0409.002 in Table \ref{spectra_summary}
using Ar~II standards from \cite{Whaling_95}, and 
iron standards taken from  Learner \& Thorne \cite{Learner_88} and
Nave et al. \cite{Nave_91}. The error bars represent the statistical uncertainty
in the measurement of the wavenumber.
\label{small_aperture}}

\end{figure}

Figure \ref{small_aperture} shows the calibration of one of our spectra
using Ar~II and iron lines from Refs. \cite{Whaling_95,Learner_88,
Nave_91} as standards. The
calibration constant k$_{\mbox{eff}}$ does not depend on 
wavenumber and is the same for all three sets of standards 
to within 1:10$^8$ when the iron lines from 
the Refs \cite{Learner_88,Nave_91} are adjusted to the wavenumber scale of
\cite{Whaling_95}. The possibility of shifts due to non-uniform
illumination of the aperture were investigated by taking a spectrum
with the 5~mm diameter image of the hollow cathode lamp offset from
the 1~mm aperture by about 2~mm. This spectrum also shows good
agreement between the Ar~II and iron calibrations.

Many of the early interferograms from the NSO FT spectrometer were
asymmetrically sampled, with a much larger number of points on one
side of  zero optical path difference than the other. A Fourier
transform of an asymmetrically-sampled interferogram gives a spectrum with a large,
antisymmetric imaginary part \cite{Learner_95}. A small error in the
phase correction causes a small part of this antisymmetric
imaginary part to be rotated into the real part of the spectrum,
distorting the line profiles and causing a wavenumber shift. 
The zero optical path difference in spectrum k19 is roughly 1/5 of the
way through the interferogram. For a Gaussian profile with a full
width at half maximum of W, this produces a wavenumber shift of roughly
0.3W per radian of phase error as shown in Fig. 3 of
\cite{Learner_95}. 

\begin{figure}
\includegraphics[width=130mm]{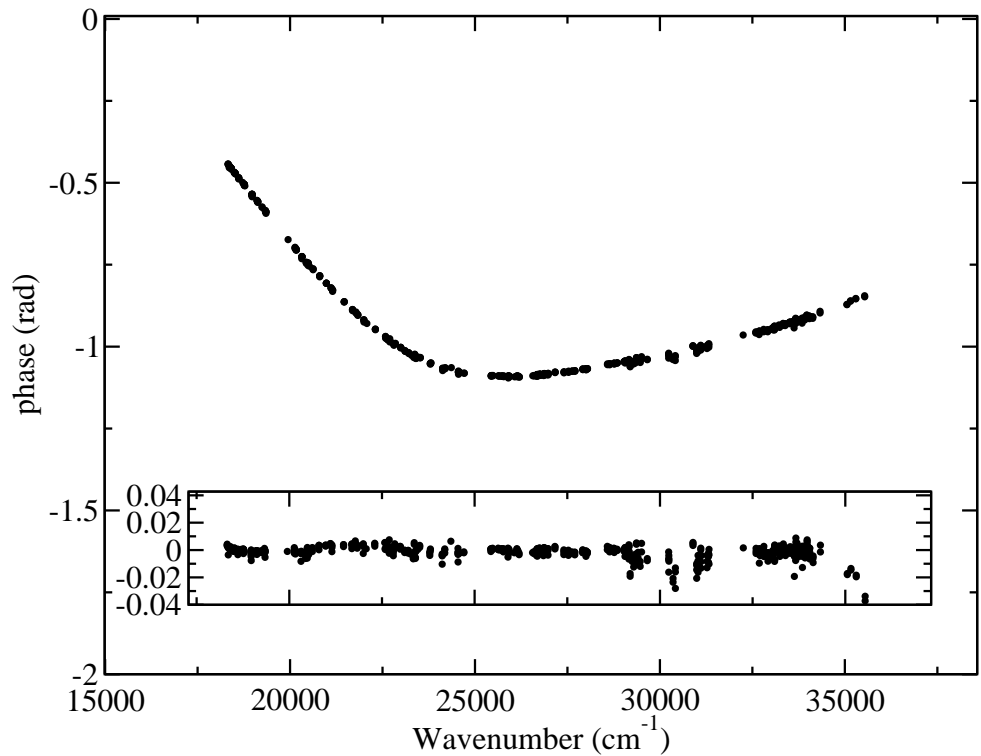}
\caption{ Phase in the master spectrum, k19, used in Learner \& Thorne
\cite{Learner_88}. The insert shows the residual phase after fitting
the points to an 11$^{\mbox{th}}$ order polynomial.
\label{k19_phase}} 
\end{figure}

\begin{figure}
\includegraphics[width=130mm]{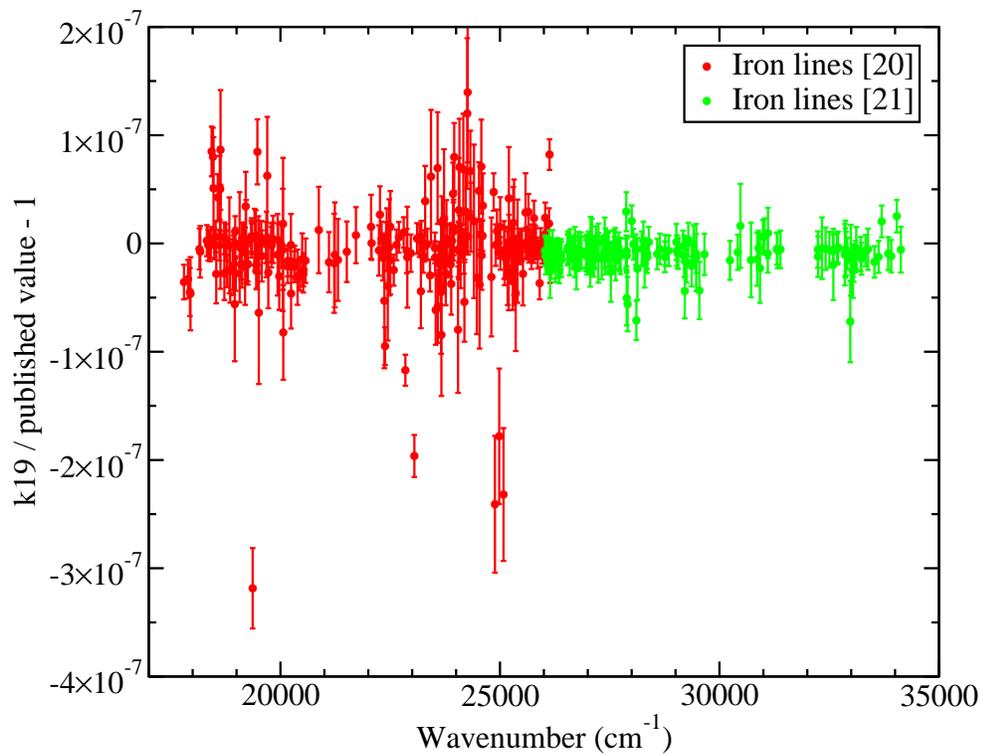}
\caption{ (Color online) Comparison of wavenumbers in the master spectrum, k19,
calibrated from Ar~II standards from \cite{Whaling_95} with iron standards taken from
\cite{Learner_88} and \cite{Nave_91} adjusted to the scale of \cite{Whaling_95}. 
The error bars represent the statistical uncertainty
in the measurement of the wavenumber.
\label{k19_wavenos}}
\end{figure}

We decided to re-examine the phase curve for the
master spectrum, k19, against which all the other iron spectra used in
\cite{Learner_88,Nave_91,Nave_97} were calibrated. The
original interferogram for this spectrum was obtained from the NSO
Digital Archives \cite{NSO_archive} and re-transformed using Xgremlin. The
phase is plotted in Fig. \ref{k19_phase}. The residual phase error
after fitting an 11$^{\rm th}$ order polynomial is less than 10~mrad
for almost all wavenumbers below 35~000~cm$^{-1}$. This corresponds to
an error of 3.6x10$^{-4}$~cm$^{-1}$ for a linewidth of 0.12~cm$^{-1}$.
Above 35~000~cm$^{-1}$ the polynomial no longer fits the points
adequately and consequently these points were not used in the
comparison. Wavenumbers were measured in the re-transformed spectrum
and calibrated with the 28 Ar~II lines recommended in
Ref. \cite{Learner_88} using the values of
Ref. \cite{Whaling_95}. Iron lines were then compared with those from
papers \cite{Learner_88,Nave_91}. The result is shown in
Fig. \ref{k19_wavenos}. The two measurements agree to within
1:10$^8$. This confirms that the original phase correction of k19 was
accurate and the wavenumbers in Ref. \cite{Learner_88} and Table 3 of
Ref. \cite{Nave_91} (2929~\AA\ to 3841~\AA ) are not affected by phase
errors.

We conclude that the wavenumbers measured in the master spectrum, k19, are
accurate. Although results from this spectrum were used
in Learner \& Thorne \cite{Learner_88} and Table 3 of Nave et
al. \cite{Nave_91}, it did not dominate the weighted average values
reported in these papers. 

\subsection{Calibration of the ultraviolet spectra\label{cal_uv}}

\begin{figure}
\includegraphics[width=130mm]{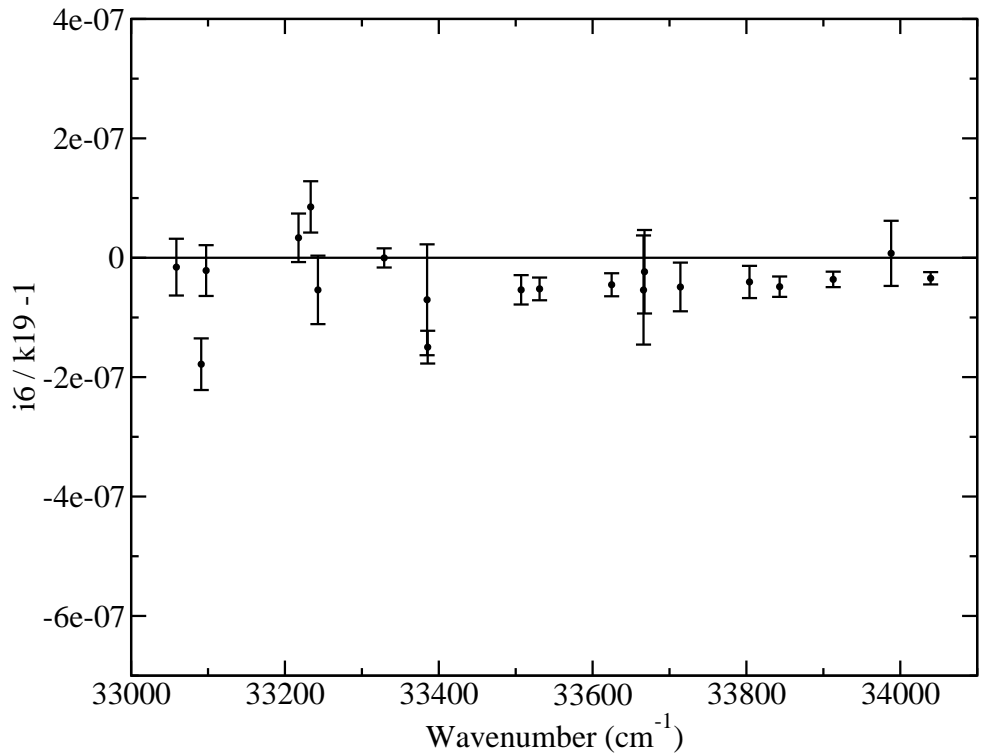}
\caption{ Comparison of wavenumbers in the master spectrum, k19, with
those in i6, the main spectrum contributing to Table 4 of  \cite{Nave_91} in this
wavelength region.
\label{k19_fe6}}
\end{figure}

Tables 4 and 5 in Nave et al. \cite{Nave_91} cover wavenumbers from
33~695~cm$^{-1}$ to 54~637~cm$^{-1}$ in Fe~I and Fe~II
respectively. The wavenumbers in these tables were measured using the
vacuum ultraviolet FT spectrometer at IC. The calibration of these
spectra was transferred from the master spectrum (k19 in Table
\ref{spectra_summary}) using a bridging spectrum (i56 in Table
\ref{spectra_summary}), as described in section \ref{cal}. The
principal spectrum covering wavenumbers below 35~000~cm$^{-1}$ in
Table 4 of \cite{Nave_91} is i6 in Table \ref{spectra_summary}. It
overlaps with the master spectrum between 33~000~cm$^{-1}$ and
34~000~cm$^{-1}$. Figure \ref{k19_fe6} shows a comparison of
wavenumbers in i6 with the master spectrum k19.  The wavenumbers in
spectrum i6 are systematically smaller than in k19 by 3.9$\pm$0.5
parts in 10$^8$. Although the region of overlap of i6
with k19 is small and is thus insensitive to non-linearities in the
wavenumber scale, this result supports our earlier speculation in Nave
\& Sansonetti \cite{Nave_04} that the calibration of the UV data using
the bridging spectrum may be incorrect.  Based on the comparison of
Fig \ref{k19_fe6}, we conclude the wavenumbers in Tables 4 and 5 of
Ref. \cite{Nave_91} should be increased by 10.6 parts in 10$^8$,
consisting of 3.9 parts in 10$^8$ to correct the transfer of the
calibration to the ultraviolet and an additional 6.7 parts in 10$^8$
to put all the spectra on the wavenumber scale of Whaling et
al. \cite{Whaling_95}.

We compared our corrected values for iron lines in the UV to results
of Aldenius et al. \cite{Aldenius_06,Aldenius_09}, who present
wavenumbers of iron lines measured in a high-current hollow cathode
lamp using a UV  FT spectrometer similar to the one used in
\cite{Nave_91}. Instead of recording a pure iron spectrum,
they included small pieces of Mg, Ti, Cr, Mn and Zn in their Fe
cathode. This ensured that spectral lines due to all of these species
were placed on the same wavenumber scale, which was calibrated using
the Ar~II lines of Whaling et al. \cite{Whaling_95}. Table
\ref{feuv_table} compares the wavenumbers of Ref. \cite{Aldenius_09}
with the corrected values of \cite{Nave_91}. Although the wavenumbers
of Ref. \cite{Aldenius_09} agree with our revised values within their
joint uncertainties, they are systematically smaller by 
3.7 parts in 10$^8$. Although this might suggest that it is incorrect to
increase the wavenumbers of Ref. \cite{Nave_91}, it might also
indicate that the wavenumbers of Ref. \cite{Aldenius_09} need to be
increased.

Fortunately, there are data that allow us to test these
alternatives. In addition to iron lines, the spectra in
Ref. \cite{Aldenius_09} contained four lines due to Mg~I and Mg~II
that have since been measured using frequency comb spectroscopy
\cite{Salumbides_06,Hannemann_06, Batteiger_09} with much higher
accuracy than achievable using FT spectroscopy. Table \ref{mguv_table}
compares the wavenumbers of these four magnesium lines from
\cite{Aldenius_09} with those derived from frequency comb measurements
of isotopically pure values. For
this comparison, the results of \cite{Aldenius_09} have been increased
by 3.7 parts in 10$^8$, as suggested by the comparison of Fe~II lines
in Table \ref{feuv_table}. With this adjustment, the results of
Aldenius agree with the frequency comb values within their joint
uncertainties, having a mean deviation of -0.7$\pm$3 parts in
10$^8$. Without the adjustment the mean deviation would be
(-4$\pm$3)$\times$10$^{-8}$. 

We conclude that the wavenumbers in Tables 4 and 5 of Nave et
al. \cite{Nave_91} should be increased by 10.6 parts in 10$^8$:
3.9 parts in 10$^8$ to correct for the incorrect transfer of the
calibration from the master spectrum to the ultraviolet and 6.7
parts in 10$^8$ to put all of the spectra on the scale of Whaling et
al. \cite{Whaling_95}. We have performed this correction in the
following sections of the current paper. The wavenumbers of Aldenius et
al. \cite{Aldenius_09} should be increased by 3.7 parts in 10$^8$ to
put them on the same scale. This adjustment of scale brings the
measurements of lines of Mg~I and Mg~II in \cite{Aldenius_09} into
agreement with the more accurate frequency comb values  
\cite{Salumbides_06,Hannemann_06,Batteiger_09}. 

\section{Wavenumbers of a$^6$D - y$^6$P transitions}\label{a6D-y6P}

The wavenumbers of the a$^6$D - y$^6$P transitions can be obtained
either from direct measurements or from energy levels derived from a
larger set of experimental data (Ritz wavenumbers). Direct
measurements will have larger uncertainties due to the cumulative
addition of the uncertainties in the transfer of the calibration from
the visible to the UV. Ritz wavenumbers are more accurate due to the
increased redundancy, but use of a large set of experimental data to
derive the energy levels makes it less clear exactly how the Ritz
wavenumbers are derived. We illustrate this process by using a small
subset of the strongest transitions that determine the y$^6$P levels
that are present in the visible and ultraviolet regions of the
spectrum where we have corrected the wavenumber calibration.

\begin{figure}
\includegraphics[width=130mm]{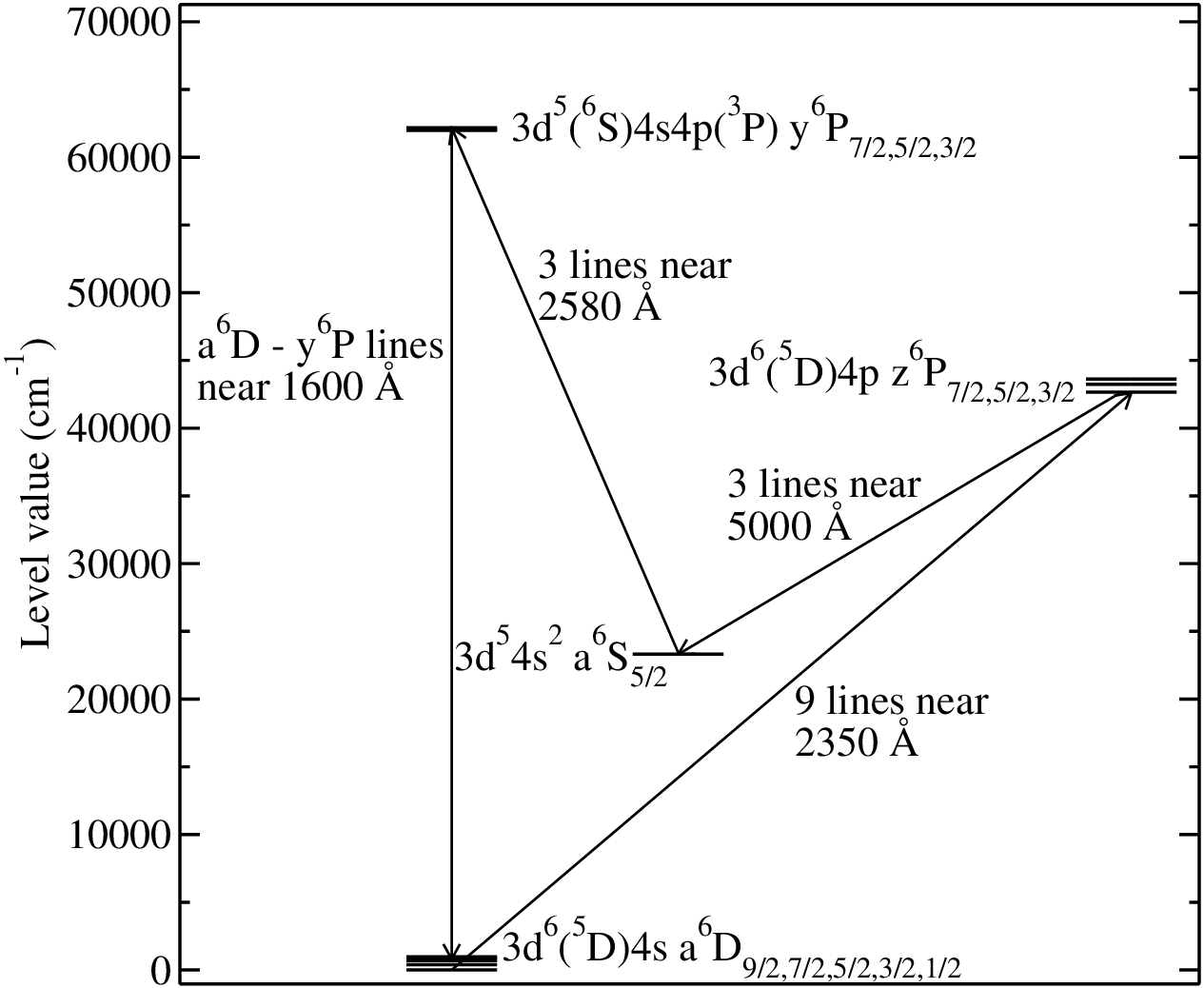}
\caption{ Partial term diagram of Fe~II showing the determination of
the y$^6$P levels using transitions in the UV and visible regions
\label{y6P_terms}}
\end{figure}

The y$^6$P levels can be determined from three sets of
lines in the UV and visible as shown in Fig. \ref{y6P_terms}. The
first set of nine lines near 2350~\AA\ determines the three $\rm
3d^6(^6D)4p\,z^6P$ levels. All nine lines are present in archival
spectra from IC which we have recalibrated using the results of
section \ref{cal_uv}. Two of the nine lines are blended with other
lines and a third, between $\rm a^6D_{9/2}$ and $\rm z^6P_{7/2}$, is
self-absorbed in the IC spectra. These lines are unsuitable for determining
the z$^6$P levels. The recalibrated values of the remaining six lines
are shown in column 4 of Table \ref{z6P_table}. Each line is observed
with a signal-to-noise ratio of over 100 in at least eight spectra,
all of which agree within 0.006~cm$^{-1}$. The wavenumbers in Table
\ref{z6P_table} are weighted mean values of the individual
measurements and the standard deviation in the last decimal place is
given in parenthesis following the wavenumber. The lower levels in
column 3 are determined from between 10 and 20 different transitions
to upper levels and have been optimized to the archival spectra with
the program {\sc lopt} \cite{lopt} (described below). The total
standard uncertainty in the upper levels includes the calibration
uncertainty of 2.3x10$^{-8}$ times the level value. 

The second set of three transitions around 5000~\AA\ determines the
3d$^5$4s$^2$\ a$^6$S$_{5/2}$ level from the three z$^6$P
levels. These lines are present in k19 and other archival spectra
taken at NSO that we have recalibrated to correspond to the
wavenumber scale of Whaling et al. \cite{Whaling_95}. Each line is
present in five spectra, all of which agree 
within 0.0035~cm$^{-1}$. Wavenumbers for these transitions are shown in Table
\ref{a6S_table} and give a mean value of
(23317.6344$\pm$0.0010)~cm$^{-1}$ for the 3d$^5$4s$^2$\ a$^6$S$_{5/2}$
level. 

Finally, the y$^6$P levels can be determined from the a$^6$S$_{5/2}$
level from three lines around 2580~\AA , present in the IC
spectra. The recalibrated wavenumbers are shown in Table \ref{y6P_table} with the
resulting y$^6$P level values. These values were used to calculate
Ritz wavenumbers for the a$^6$D - y$^6$P transitions, as shown in the
third column of Table \ref{y6p_ritz}.

Alternate values for the Ritz wavenumbers of the a$^6$D-y$^6$P
transitions can be obtained from energy levels optimized using
wavenumbers from the archival Fe~II spectra from NSO and IC corrected
according to sections \ref{vis_NIST} and \ref{cal_uv}. The program
{\sc lopt} \cite{lopt} was used to derive optimized values for 939 energy levels
from 9567 transitions. Weights were assigned proportional to the
inverse of the estimated variance of the wavenumber. Lines 
with more than one possible classification, lines that were blended,
or for which the identification was questionable were assigned a low
weight. Two iterations were made. In the first, lines connecting the
lowest a$^6$D term to higher $\rm 3d^6\,(^5D)4p$ levels were
assigned a weight proportional to the inverse of the statistical
variance of the wavenumber, omitting the calibration uncertainty. This
was done to obtain accurate values and uncertainties for the a$^6$D
intervals. These intervals are determined from differences between
lines close to one another in the same spectrum sharing the same
calibration factor. Hence the calibration uncertainty does not
contribute to the uncertainty in the relative values of these energy
levels. The values of the a$^6$D levels obtained in this iteration are
given in column 3 of table \ref{z6P_table}. In the second iteration,
the a$^6$D levels were fixed to the values and uncertainties
determined from the first iteration. The weights of the $\rm a^6D
- 3d^6\,(^5D)4p$ transitions were assigned by combining in quadrature
the statistical uncertainty in the measurement of the line position
and the calibration uncertainty in order to obtain
accurate uncertainties for the $\rm 3d^6\,(^5D)4p$ and higher
levels. The values of the y$^6$P levels are given in column 4 of table
\ref{y6P_table}. Ritz wavenumbers for the $\rm a^6D -
3d^6\,(^5D)4p$ transitions based on these globally optimized level
values are presented in column 5 of table \ref{y6p_ritz}. 

The corrected experimental wavenumbers from the
archival spectra are given in column 4 of table \ref{y6p_ritz}. 
The main contribution to the uncertainty in the experimental
wavenumbers is from the calibration and consists of two components --
the uncertainty in the standards and the uncertainty in calibrating
the spectrum. The calibration uncertainty is common to all lines in the
calibrated spectrum and hence must be added to the uncertainties of
wavenumbers measured using transfer standards, rather than added in
quadrature as would be the case for random errors. Hence the
uncertainty in the wavenumbers increases with each calibration step,
resulting in larger uncertainties at the shortest wavenumbers which
are furthest from the calibration standards. The experimental standard
uncertainties in Table \ref{y6p_ritz} are determined by combining in
quadrature the statistical uncertainty in determining the line
position and the calibration uncertainty of 4$\times$10$^{-8}$ times
the wavenumber. The experimental wavenumber and both
Ritz wavenumbers agree within their joint uncertainties. The Ritz
wavenumbers determined from optimized energy levels have the smallest
uncertainties. Wavelengths corresponding to these wavenumbers are
given in column 7. 

\section{A re-examination of Fe~II wavenumbers from Nave \&
Sansonetti\cite{Nave_04}}\label{ge_comp} 

The Fe~I lines in Nave \& Sansonetti \cite{Nave_04} were calibrated
with respect to lines of Ge~I. Figure 3 of that paper showed that the
calibration factor k$_{\mbox{eff}}$ derived from Fe~I and Fe~II lines is smaller than that
derived from Ge~I by 6.5 parts in 10$^8$. We attributed this to a
possible problem in the transfer of the wavenumber calibration of the
Fe~I and Fe~II lines from the region of the Ar~II wavenumber standards
to the vacuum ultraviolet, thus suggesting that the wavenumber
standards in \cite{Nave_91} are too small. In section \ref{cal_uv} we have
confirmed that the wavenumbers in Tables 4 and 5 of Nave et al.
\cite{Nave_91} should be increased by 3.9 parts in 10$^{8}$ due to the
transfer of the calibration. This reduces but does not fully explain
the calibration discrepancy in \cite{Nave_04}.

\begin{figure}
\includegraphics[width=130mm]{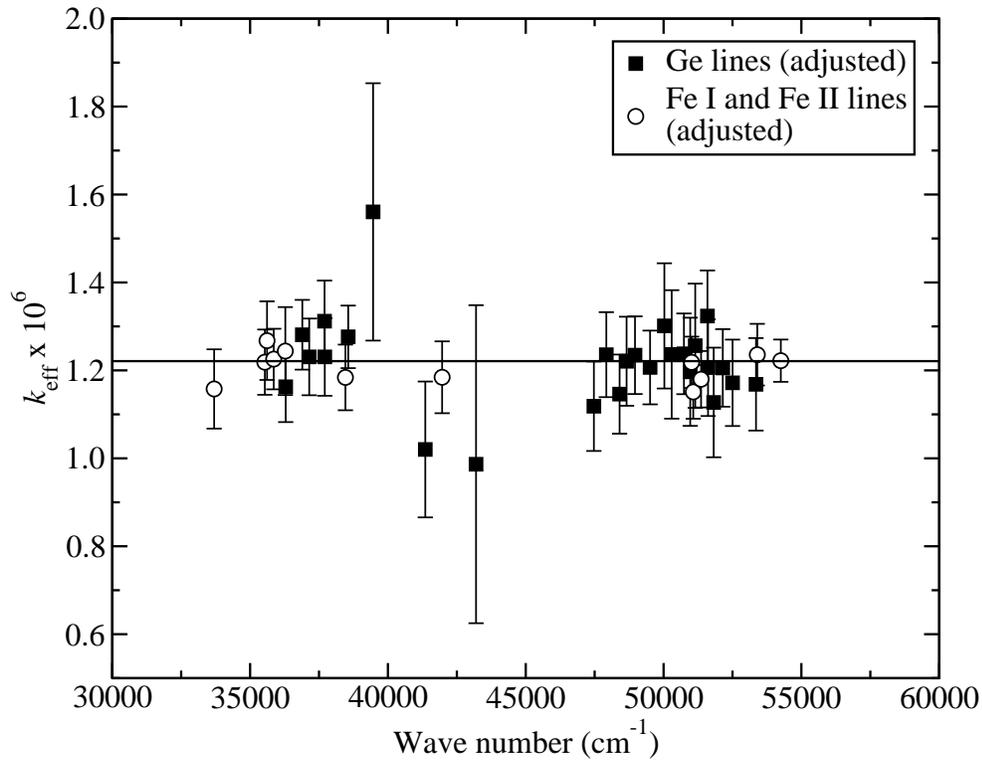}
\caption{
Figure 3 from Nave \& Sansonetti \cite{Nave_04}, with all of the Ge~I wavenumbers
reduced by 1.4 parts in 10$^8$ and the Fe~I and Fe~II wavenumbers increased
by 3.9 parts in 10$^8$. The mean value of k$_{\mbox{eff}}$ for the
Ge~I wavenumbers is $(1.221 \pm 0.020) \times$10$^{-6}$, in agreement
within the joint uncertainties with the value of $(1.206 \pm 0.020)
\times$10$^{-6}$ from the Fe~I and Fe~II lines
\label{paperVI_fig3} }
\end{figure}

The Ge~I lines used to calibrate the spectra in Ref. \cite{Nave_04}
were measured by Kaufman and Andrew \cite{Kaufman_62}. The wavenumber
standard they used was the 5462~\AA\ line of $^{198}$Hg emitted
by an electrodeless discharge lamp maintained at a temperature of
$19\,^{\circ}\rm{C}$, containing Ar at a pressure of 400~Pa (3
Torr). The vacuum wavelength of this line was
assumed to be 5462.27075~\AA. This value was based on a vacuum
wavelength of 5462.27063~\AA\ measured in the same lamp at
$7\,^{\circ}\rm{C}$ \cite{Kaufman_62A}, with an adjustment for the
different temperature using the measurements of Emara
\cite{emara}. The 5462~\AA\ line was remeasured by Salit et al. \cite{Salit_04}
using a temperature of $8\,^{\circ}\rm{C}$. A value of
(5462.27085$\pm$0.00007)~\AA\ was obtained. More recent work by Sansonetti \&
Veza \cite{Sansonetti_10} gives the wavelength of this line 
as 5462.270825(11)~\AA, in agreement with
\cite{Salit_04} but more precise. Adoption of this value for the
wavelength of the $^{198}$Hg line implies that all of the Ge~I
wavenumbers in \cite{Kaufman_62} should be decreased by 1.4 parts in
10$^8$. Figure \ref{paperVI_fig3} shows how Fig. 3 in \cite{Nave_04}
(Spectrum lp0301 in Table \ref{spectra_summary})
changes with the adjustment of both the iron and germanium
wavenumbers. The calibrations based on Ge and Fe lines now
differ by only 1.5 parts in 10$^8$, which is within the joint uncertainties.
We thus conclude that the calibration derived from Fe~I and Fe~II
lines is in agreement with that derived from Ge~I when both sets of
standards are adjusted to correspond with the most recent
measurements.

Spectrum lp0301 in Table \ref{spectra_summary} can be used to
calibrate  spectrum fe1115 in Table \ref{spectra_summary} 
referred to in the last paragraph of section \ref{previous}. A value of
(62171.634$\pm$0.006)~cm$^{-1}$ is obtained for the wavenumber of the 
$\rm a^6D_{9/2} - y^6P_{7/2} $ line, corresponding to a wavelength of
(1608.45057$\pm$0.00016)~\AA, This disagrees with the Ritz value by
1.7 times the joint uncertainty and marginally disagrees with the
experimental values of Table \ref{y6p_ritz}. The mean difference in  
the experimental values for all nine $\rm a^6D - y^6P $ lines is
(0.008$\pm$0.004)~cm$^{-1}$. We believe this difference is due to a small
slope in the calibration of lp0301 but have been unable to confirm this
with our data. The principal contributors to the
uncertainty are the uncertainty in the iron and germanium standards,
the uncertainty in calibrating the spectrum in ref. \cite{Nave_04}
from these standards, and the uncertainty in calibrating 
spectrum fe1115 from spectrum lp0301.

\section{Conclusions}

We investigated the wavenumber scale of published Fe~I and Fe~II
lines using new spectra recorded with the NIST 2-m FT spectrometer and
a re-analysis of archival spectra. Our new
spectra confirm the wavenumber scale of visible-region iron lines
calibrated using the Ar~II wavenumber standards of Whaling et
al. \cite{Whaling_95}. 

Having confirmed the wavenumber scale of iron lines in the visible and
ultraviolet regions, we have used lines from these spectra to derive Ritz 
values for the wavenumbers and wavelengths of lines in the
$\rm a^6D - y^6P$ multiplet of Fe~II (UV 8). Ritz wavenumbers derived
using two different methods agree with one another and with directly
measured wavenumbers within the joint uncertainties. We recommend a
value of 1608.45081$\pm$0.00007~\AA\ for the wavelength of the $\rm
a^6D_{9/2} - y^6P_{7/2}$ line of Fe~II, which is an important line for
detection of changes in the fine-structure constant during the
history of the Universe using quasar absorption-line spectra. 

We find that the wavenumbers in Learner \& Thorne \cite{Learner_88}
and Table 3 of Nave et al. \cite{Nave_91} should be increased by 6.7
parts in 10$^8$ to put them on the scale of the Ar~II lines of Whaling
et al. \cite{Whaling_95}. The wavenumbers in Tables 4 and 5 of
Ref. \cite{Nave_91} should be increased by 10.6 parts in 10$^8$ to put
them on the Ar~II scale of Ref. \cite{Whaling_95} and to correct for
an error in the transfer of this wavenumber scale to the ultraviolet.
The Ge~I wavenumbers of Kaufman \& Andrew \cite{Kaufman_62} and all
the wavenumbers in Nave \& Sansonetti \cite{Nave_04} should be
decreased by 1.4 parts in 10$^8$ to put them on the scale of recent
measurements of the $^{198}$Hg line at 5462~\AA.

\section{Acknowledgments}

We thank Michael T. Murphy for alerting us to the importance of the
Fe~II line at 1608~\AA . We also thank Anne P. Thorne and Juliet
C. Pickering for helpful discussions on the calibration of FT
spectrometers and the linearity of the FT wavenumber scale. This work
is partially supported by NASA inter-agency agreement NNH10AN38I.

\clearpage

\clearpage
\begin{table}
\caption{Proposed corrections to previous papers.
\label{corrections}
}
\begin{minipage}{8in}
\begin{footnotesize}
\begin{tabular}{llllll}
\hline
Reference                       & Wavenumber     & Previous              & Previous    & New                                        & Correction to             \\
                                & range          & standard              & uncertainty & standard                                   & wavenumber scale          \\
                                & (cm$^{-1}$)    &                       & (cm$^{-1}$) &                                            &                           \\
\hline
\cite{Learner_88}               & 17350 - 26140  & Ar~II~\cite{Norlen_73}      & 0.001 & Ar~II~\cite{Whaling_95}                    &  (+6.7$\pm$1.8)x10$^{-8}$ \\
Table 3 of \cite{Nave_91}       & 26027 - 34131  & Ar~II~\cite{Norlen_73}, i56 & 0.002 & Ar~II~\cite{Whaling_95}                    &  (+6.7$\pm$1.8)x10$^{-8}$ \\
Tables 4 \& 5 of \cite{Nave_91} & 33695 - 54637  & Ar~II~\cite{Norlen_73}, i56 & 0.002 & Ar~II~\cite{Whaling_95}, Fig.\ref{k19_fe6} & (+10.6$\pm$2.3)x10$^{-8}$ \\
\cite{Nave_97}                  & 50128 - 107887 & Ar~II~\cite{Norlen_73}, i56 & 0.005 & Ar~II~\cite{Whaling_95}, Fig.\ref{k19_fe6} & (+10.6$\pm$2.3)x10$^{-8}$ \\
\cite{Nave_04}                  & 51613 - 51692  & Ge~I~\cite{Kaufman_62}      & 0.002 & $^{198}$Hg \cite{Sansonetti_10}            & (-1.4$\pm$2)x10$^{-8}$    \\
\cite{Aldenius_09}              & 38458 - 44233  & Ar~II~\cite{Whaling_95}     & 0.002 & Table \ref{feuv_table},\ref{mguv_table}    & (+3.7$\pm$3)x10$^{-8}$    \\
\cite{Kaufman_62} (Ge~I \& Ge~II)& 8283 - 100090 &
                                           $^{198}$Hg,\cite{Kaufman_62A,emara}      & $<$0.006 & $^{198}$Hg \cite{Sansonetti_10} & (-1.4$\pm$1.8)x10$^{-8}$  \\
\cite{Norlen_73} (Ar~II)$^a$
                              & 4348 - 5145   & $^{86}$Kr Engelhard lamp & $<$0.001 & Ar~II~\cite{Whaling_95}                    & (+6.7$\pm$0.8)x10$^{-8}$  \\
\hline
\end{tabular}
\footnotetext[1]{The proposed correction has only been confirmed for the 28 Ar~II lines recommended in \cite{Learner_88}.}
\end{footnotesize}
\end{minipage}
\end{table}

\clearpage
\begin{table}
 \caption{Summary of spectra
\label{spectra_summary}}.
\begin{tabular}{llllll}
\hline
Spectrum  & Instrument & Date  & Wavelength  & Calibration & Comments \\
          &            & y/m/d & Range  (\AA)            & Spectrum \\ 
\hline
k19 \cite{Nave_91} & NSO  & 81/07/22  & 2800 to 5600 & Ar II\cite{Whaling_95}  & 810622R0.009 (NSO\cite{NSO_archive}) \\
                   &      &           &             &        & A1 in \cite{Learner_88} \\
i56 \cite{Nave_91} & IC   &           & 2270 to 4170 & k19    & \\
i6 \cite{Nave_91}  & IC   & 89/11/07  & 2220 to 3030 & i56    & \\
lp0301         & NIST VUV & 02/03/01  & 1830 to 3194 & Ge~I,II\cite{Kaufman_62} & Figs. 3 \&4 in \cite{Nave_04} \\
fe1115         & NIST VUV & 02/11/15  & 1558 to 2689 & lp0301 & \\
fe0409.002     & NIST 2-m & 09/04/09  & 2748 to 5765 & Ar~II \cite{Whaling_95}, \\
               &          &           &             & Fe~I,II\cite{Learner_88,Nave_91} \\
\hline 
\end{tabular}
\end{table}

\clearpage
\begin{table}
 \caption{Comparison of wavenumbers of Fe lines in Aldenius et
   al. \cite{Aldenius_09} and adjusted wavenumbers of Nave et
   al. \cite{Nave_91}. The wavenumbers of \cite{Nave_91} have been
   increased by 10.6 parts in 10$^8$. The standard uncertainty in the last
   digits of the wavenumbers and of the levels is given in parenthesis
   and is dominated by the calibration uncertainty.
\label{feuv_table}}.
\begin{tabular}{llll}
\hline
Species  & Nave et al. (cm$^{-1}$)& Aldenius et al. (cm$^{-1}$) &  (column 3 / column 2) - 1  \\
         & \cite{Nave_91}        & \cite{Aldenius_09}      &                           \\
\hline	                         
Fe~II    & 38458.9912(20)        & 38458.9908(20)          &  -1.0x10$^{-8}$           \\
Fe~II    & 38660.0535(20)        & 38660.0523(20)          &  -3.1x10$^{-8}$           \\
Fe~II    & 41968.0687(20)        & 41968.0654(20)          &  -7.7x10$^{-8}$           \\ 
Fe~II    & 42114.8374(20)        & 42114.8365(20)          &  -2.1x10$^{-8}$           \\
Fe~II    & 42658.2449(20)        & 42658.2430(20)          &  -4.5x10$^{-8}$           \\
         &                       & Mean                    &  (-3.7$\pm$2.6)x10$^{-8}$ \\
\hline 
\end{tabular}
\end{table}

\clearpage
\begin{table}
\caption{Comparison of adjusted wavenumbers of Mg lines in Aldenius et al.
\cite{Aldenius_09} with frequency comb measurements (col. 3) taken from paper
listed in the reference column. The wavenumbers from
\cite{Aldenius_09} have been increased by 3.7 parts in 10$^8$. The
standard uncertainties in the last digits of the wavenumbers are given
in parentheses.\label{mguv_table}}  
\begin{tabular}{lllll}
\hline
Species & Aldenius (cm$^{-1}$) & Frequency comb  (cm$^{-1}$)   
                                          & (column 2/column 3) -1 & Reference \\
      & \cite{Aldenius_09}                 \\
\hline
Mg~I  & 35051.2817(20)   & 35051.2808(2)  & 2.6x10$^{-8}$   & \cite{Salumbides_06} \\
Mg~II & 35669.3039(20)   & 35669.30440(6) & -1.3x10$^{-8}$  & \cite{Batteiger_09} \\
Mg~II & 35760.8523(20)   & 35760.85414(6) & -5.1x10$^{-8}$  & \cite{Batteiger_09} \\
Mg~I  & 49346.7730(30)   & 49346.77252(7) & 1.0x10$^{-8}$   & \cite{Hannemann_06} \\
      &                  & Mean           & (-0.7$\pm$3)x10$^{-8}$ \\
\hline
\end{tabular}
\end{table}

\clearpage
\begin{table}
\caption{Determination of the z$^6$P levels of Fe~II from transitions
to the ground term around 2350~\AA . The statistical uncertainty in
the last decimal place of the wavenumbers is given in parenthesis. The
total standard uncertainty includes the uncertainty in the calibration.
\label{z6P_table}
}
\begin{tabular}{lllll}
\hline
Lower      & Upper   & lower level value & Wavenumber & Upper level value \\
level      & level   & cm$^{-1}$         & cm$^{-1}$  & cm$^{-1}$ \\
\hline
a$^6D_{5/2}$ & z$^6P_{7/2}$ & 667.6829(5) & 41990.5610(3) & 42658.2439(6) \\
a$^6D_{7/2}$ & z$^6P_{7/2}$ & 384.7872(4) & 42273.4573(4) & 42658.2445(6) \\
      &   &   & Mean      & 42658.2442(5) \\
      &   &   & Total uncertainty    &     0.0011 \\
\\
a$^6D_{5/2}$ & z$^6P_{5/2}$ & 667.6829(5) & 42570.9226(4) & 43238.6055(6) \\
a$^6D_{7/2}$ & z$^6P_{5/2}$ & 384.7872(4) & 42853.8188(4) & 43238.6060(6) \\
            & & & Mean   & 43238.6058(5)  \\
      &   &   & Total uncertainty    &     0.0011  \\
\\
a$^6D_{1/2}$ & z$^6P_{3/2}$ & 977.0498(6) & 42643.9332(4) & 43620.9830(7) \\
a$^6D_{5/2}$ & z$^6P_{3/2}$ & 667.6829(5) & 42953.2994(5) & 43620.9823(7) \\
            & & & Mean & 43620.9827(6) \\
      &   &   & Total uncertainty    &     0.0012  \\
\hline
\end{tabular}
\end{table}

\clearpage
\begin{table}
\caption{Determination of the a$^6$S$_{5/2}$ level of Fe~II using
transitions from the z$^6$P levels. The statistical uncertainties in the last digits
of the wavenumber and levels are given in parentheses. The total standard
uncertainty of the a$^6S$ level includes a contribution of
4$\times10^{-8}$ times the level uncertainty due to the calibration.
\label{a6S_table}} 
\begin{tabular}{llll}
\hline
Upper level & Upper level value  & Wavenumber     & a$^6$S level  \\
            & cm$^{-1}$          & cm$^{-1}$      & cm$^{-1}$     \\
\hline
z$\rm ^6P_{7/2}$ & 42658.2442(5) & 19340.6092(2)  & 23317.6350(5) \\
z$\rm ^6P_{5/2}$ & 43238.6058(5) & 19920.9733(2)  & 23317.6325(5) \\
z$\rm ^6P_{3/2}$ & 43620.9827(6) & 20303.3477(3)  & 23317.6350(7) \\
                 &               & Weighted mean  & 23317.6340(3) \\
                 &               & Total uncertainty  & 0.0010 \\  
\hline
\end{tabular}
\end{table}

\clearpage
\begin{table}
\caption{Determination of the y$^6$P levels of Fe~II from
transitions to the a$^6$S level. The statistical uncertainty in the
last digits of the wavenumbers and levels is given in
parentheses. The total standard uncertainty is common to all levels and
includes a contribution of 4$\times10^{-8}\sigma$ due to the
calibration. The last column contains the level value and standard
uncertainty in parenthesis with respect to the ground level obtained
from the {\sc lopt} program as described in section \ref{a6D-y6P}
\label{y6P_table}. 
}
\begin{tabular}{llll}
\hline
Upper level      & Wavenumber     & Upper level value & Level value
from {\sc lopt} \\
                 & cm$^{-1}$      & cm$^{-1}$         & cm$^{-1}$    \\
\hline
y$\rm ^6P_{3/2}$ & 38657.2997(14) & 61974.9347(14)    & 61974.9325(24)   \\
y$\rm ^6P_{5/2}$ & 38731.4041(7)  & 62049.0381(8)     & 62049.0408(27)   \\
y$\rm ^6P_{7/2}$ & 38853.9885(4)  & 62171.6225(5)     & 62171.6245(27)   \\
            & Total uncertainty  & 0.003              &   \\
\hline
\end{tabular}
\end{table}

\clearpage
\begin{table}
\caption{Experimental and Ritz wavenumbers for the a$^6$D-y$^6$P
multiplet. The standard uncertainties in the last digits of the
wavenumbers and wavelengths are given in parentheses.\label{y6p_ritz}}
\begin{minipage}{8in}
\begin{tabular}{llllll}
\hline
Lower      & Upper & $\sigma_{a6S}^a$ 
&$\sigma_{exp} ^b$ 
&$\sigma_{Ritz} ^c $ 
& $\lambda_{Ritz} ^d $ 
\\
level          & level          & cm$^{-1}$    & cm$^{-1}$     & cm$^{-1}$    & \AA \\
\hline					        		                    
a$^6$D$_{1/2}$ & y$^6$P$_{3/2}$ & 60997.884(3) & 60997.882(3)  & 60997.8827(25) & 1639.40117(7) \\
a$^6$D$_{3/2}$ & y$^6$P$_{3/2}$ & 61112.322(3) & 61112.321(3)  & 61112.3207(25) & 1636.33125(7) \\
a$^6$D$_{5/2}$ & y$^6$P$_{3/2}$ & 61307.251(3) & 61307.247(3)  & 61307.2496(25) & 1631.12847(7) \\
a$^6$D$_{3/2}$ & y$^6$P$_{5/2}$ & 61186.426(3) & 61186.432(4)  & 61186.4290(28) & 1634.34934(7) \\
a$^6$D$_{5/2}$ & y$^6$P$_{5/2}$ & 61381.355(3) & 61381.358(3)  & 61381.3579(28) & 1629.15914(7) \\
a$^6$D$_{7/2}$ & y$^6$P$_{5/2}$ & 61664.251(3) & 61664.255(3)  & 61664.2536(27) & 1621.68508(7) \\
a$^6$D$_{5/2}$ & y$^6$P$_{7/2}$ & 61503.940(3) & 61503.945(8)  & 61503.9416(27) & 1625.91205(7) \\
a$^6$D$_{7/2}$ & y$^6$P$_{7/2}$ & 61786.835(3) & 61786.837(3)  & 61786.8373(27) & 1618.46769(7) \\
a$^6$D$_{9/2}$ & y$^6$P$_{7/2}$ & 62171.623(3) & 62171.626(4)  & 62171.6245(27) & 1608.45081(7) \\
\hline
\end{tabular}
\footnotetext[1]{Wavenumber calculated using a$^6$S as an intermediate
level} 
\footnotetext[2]{Experimental wavenumber from archival spectra
corrected according to section \ref{cal_uv}.} 
\footnotetext[3]{Ritz wavenumber calculated from all optimized energy levels} 
\footnotetext[4]{Wavelength calculated from the Ritz wavenumber in
column 5.}
\end{minipage}
\end{table}

\end{document}